\documentclass{article}
\usepackage[utf8]{inputenc}
\usepackage[a4paper, margin=25mm]{geometry}
\usepackage{authblk}
\usepackage{graphicx}
\usepackage{dcolumn}
\usepackage{bm}
\usepackage{siunitx}
\DeclareUnicodeCharacter{2212}{-}

\title{Improving the resolution of comb-based frequency measurements\\ using a  track and hold amplifier}

\author[1,*]{Matias Risaro}
\author[2]{Paolo Savio}
\author[1]{Marco Pizzocaro}
\author[1]{Filippo Levi}
\author[1]{Davide Calonico}
\author[1]{Cecilia Clivati}
\affil[1]{INRIM, strada delle cacce 91, 10135 Turin, Italy.}
\affil[2]{Fondazione LINKS, via P. C. Boggio 61, 10138 Turin, Italy.}
\affil[*]{m.risaro@inrim.it}
\date{May 2022}

\begin{document}

\maketitle

\begin{abstract}
The advent of optical frequency standards with ultimate uncertainties in the low \SI{1e-18}{} requires femtosecond frequency combs to support a similar level of resolution in the spectral transfer and the computation of optical frequency ratios. The related experimental challenges grow together with the number of optical frequencies to be measured simultaneously, as in many cases the comb's optical power does not allow reliable beatnote counting or tracking in all the spectral regions of interest. Here we describe the use of a track-and-hold amplifier to implement the gated detection, a previously proposed technique for improving the signal-to-noise ratio of the beatnote between a low-power tooth of the frequency comb and a continuous wave laser.  We  demonstrate a \SI{12}{dB} improvement in the signal-to-noise ratio of beatnotes involving a broadband-spanning optical comb as compared to traditional detection schemes. Our approach enables reliable and cycle-slip-free spectral purity transfer and reduces the system sensitivity to power drops in the comb spectrum. Being based on a single chip, it is robust, versatile and easily embedded in more complex experimental schemes. 
\end{abstract}

\section*{\label{sec:introduction}Introduction}
Since their introduction at the beginning of the century, frequency combs have revolutionized the time and frequency metrology  \cite{Diddams2020,Fortier2019}. Their broad spectrum composed of equidistant narrow lines permits a direct link between radio  and optical frequencies  \cite{Nakamura2020} and between optical frequencies, enabling to copy the spectral purity of a master laser to a different region of the spectrum and compare different species of optical clocks \cite{Inaba:13,Nicolodi2014,Rosenband2008,Nemitz2016, Beloy2021}. The most robust and widespread technology is based on Erbium-doped fiber lasers, natively covering the \SI{1550}{nm} spectrum, which is extended to other portions of the near-IR  up to the visible using nonlinear conversion techniques based on second-harmonic generation and four-wave-mixing processes in highly nonlinear fibers.

In the latest years, the unprecedented stability and accuracy of optical clocks  and the possibility to compute optical frequency ratios at the $10^{-18}$ level of uncertainty \cite{Nicholson2015,hinkley2013,Sanner2019,Ushijima2015,Beloy2021} led to more stringent requirements for the frequency combs, up to the point where the uncontrolled optical path changes in comb branches covering different portions of the spectrum were no longer sustainable, leading to instabilities as high as $10^{-16}$. Complex real-time and post-processing approaches have been developed to tackle such an issue \cite{Rolland:18,Giunta2019}. However, in most applications, the single-branch operation of the comb is preferred, where a single comb output is spread over a broad spectrum, enabling beatnotes with continuous-wave (cw) lasers at different wavelengths to be simultaneously obtained \cite{Leopardi:17,Ohmae_2017}. 
As a drawback, the available comb power per tooth is reduced and achieving adequate beatnote signal-to-noise ratio (SNR) at all wavelengths may become challenging. In turns, phase-slips in the beat counting may occur, whose rate grows exponentially with the SNR degradation \cite{proakis1995,sinclair}. The relevance of this issue in high-precision optical frequency measurements is apparent considering that even a single cycle lost in one hour may introduce a  frequency bias of the order of $10^{-18}$, if undetected. So, this kind of measurements always include tracking-oscillators to reduce the probability of cycle slips by a better filtering \cite{Johnson_2015} and/or flagging systems based on redundant counting \cite{Udem1998}. Both approaches however increase the setup complexity and may ultimately contribute additional impairments on the long-term \cite{benkler2019}.  In addition, the broadband noise due to limited SNR introduces aliasing in the measurements when the beatnotes are sampled with frequency counters.

In 2013, the gated detection of comb-to-laser beatnotes was proposed to improve the SNR exploiting  the pulsed nature of the comb \cite{Deschnes2013}. In most experimental setups, the SNR is limited by a combination of the cw laser shot noise and the electronic noise of the photodiode \cite{Reichert1999}. However, 
whilst the cw laser shot noise and the photodiode noise are mostly homogeneous over time \cite{Quinlan2013}, the  beatnote between a cw laser and the comb has a pulse-like nature. Gating the optical beatnote synchronously with the comb pulses thus enables to properly reconstruct it without losing relevant information while most of the noise, namely the fluctuations happening outside the measurement windows, are rejected. This gated approach has been implemented experimentally \cite{Deschnes2013} using bulk mixers, an additional  beatnote detection unit and cable delay-lines.

In this work, we describe an implementation of the gated detection technique that uses an off-the-shelf track and hold amplifier (THA) as the gating device and enables the real-time counting of optical beatnotes at improved SNR in a compact and versatile setup, 
 easily integrated into the conventional beatnote detection chains.

We characterize our system  in the context of spectral purity transfer between optical frequencies in a single-branch comb setup, evaluating the achievable SNR improvement and contributed uncertainty in terms of instability and cycle slips, which are key aspects in high-precision optical frequency measurements.

\section*{\label{sec:experimental_setup}Experimental setup}
\begin{figure*}
\centering
\includegraphics[width=0.9\linewidth]{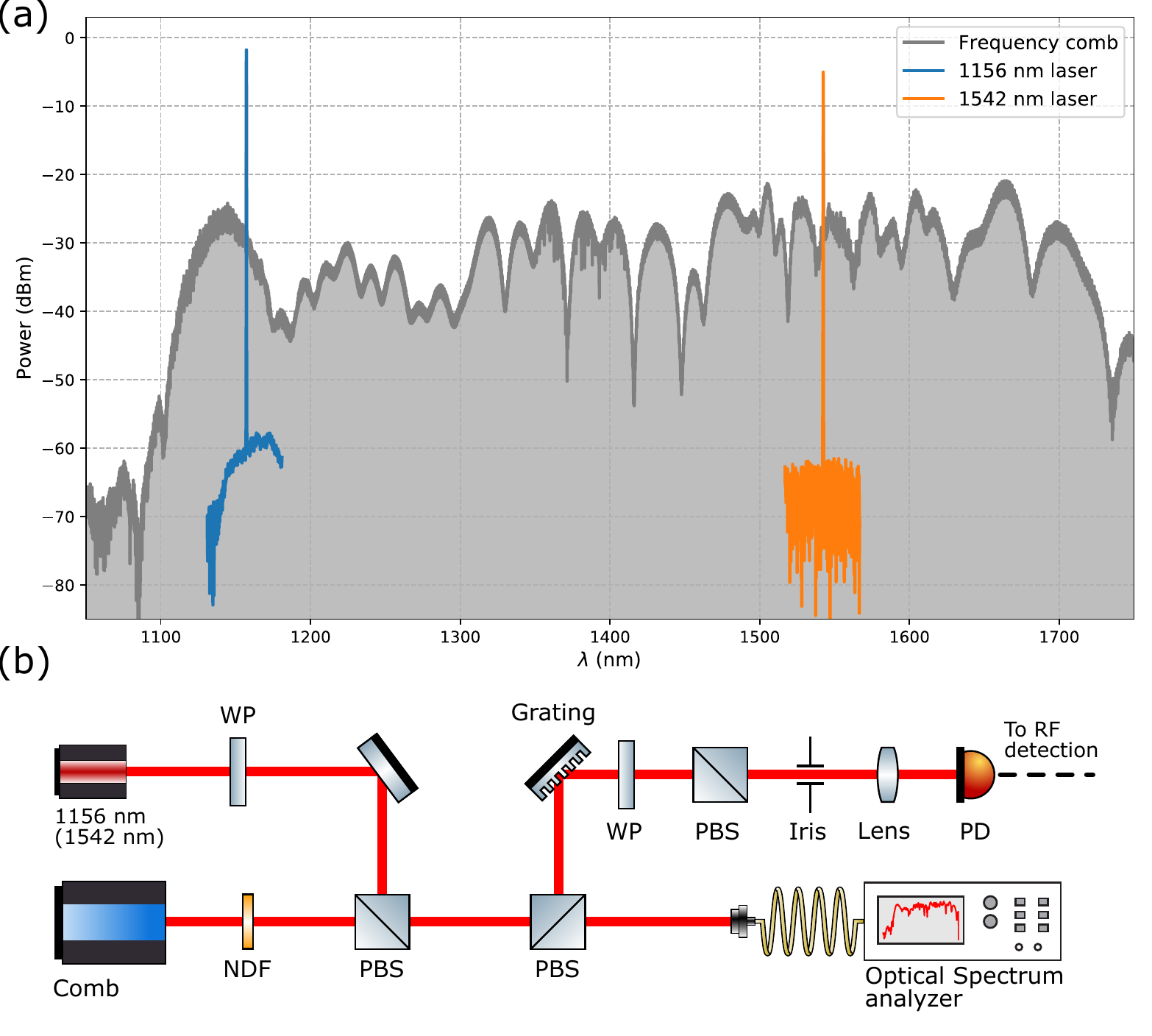}
\caption{(a) Optical spectrum of the Erbium fiber frequency comb together with the 1156 nm cavity-stabilized laser and the 1542 nm  laser. The spectra were recorded using an optical spectrum analyzer with a resolution bandwidth of 0.05 nm. (b) Simplified scheme of the free space interferometer to generate the beatnote between the frequency comb and the \SI{1156}{nm} laser. Part of the light is coupled to an optical spectrum analyzer, while the other is detected with a photodiode. PD: Photodiode; PBS: polarization beam splitter; WP: half-waveplate; NDF: neutral density filter.}
\label{fig:optic_setup}
\end{figure*}

\begin{figure}
    \centering
    \includegraphics[width=0.8\columnwidth]{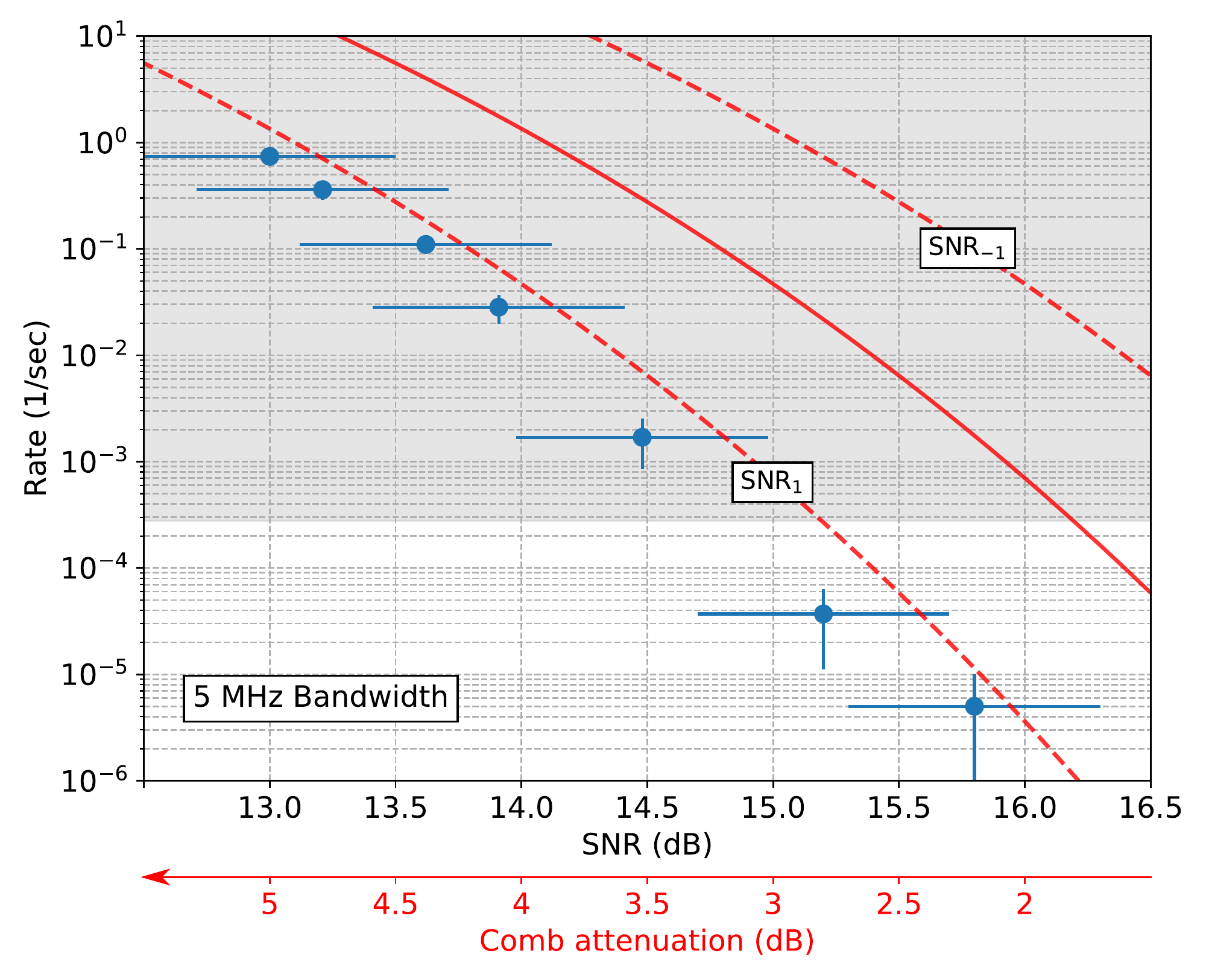}
    \caption{The rate of cycle slips versus SNR of the beatnote signal in a bandwidth of \SI{5}{MHz}. The red solid curve represents the predicted behaviour, while the dashed curves are produced by off-setting the SNR by  $+1$ and $-1$ dB. Measurements were made by attenuating the optical power of the  frequency comb as indicated by the red arrow (bottom axis). The gray area indicates the region with cycle-slips rate higher than 1/hour, which contributes a relative shift of \SI{1e-18}{} on the \SI{1156}{nm} laser counting.}
    \label{fig:snr_cycle}
\end{figure}

\begin{figure}
    \centering\includegraphics[width=0.8\columnwidth]{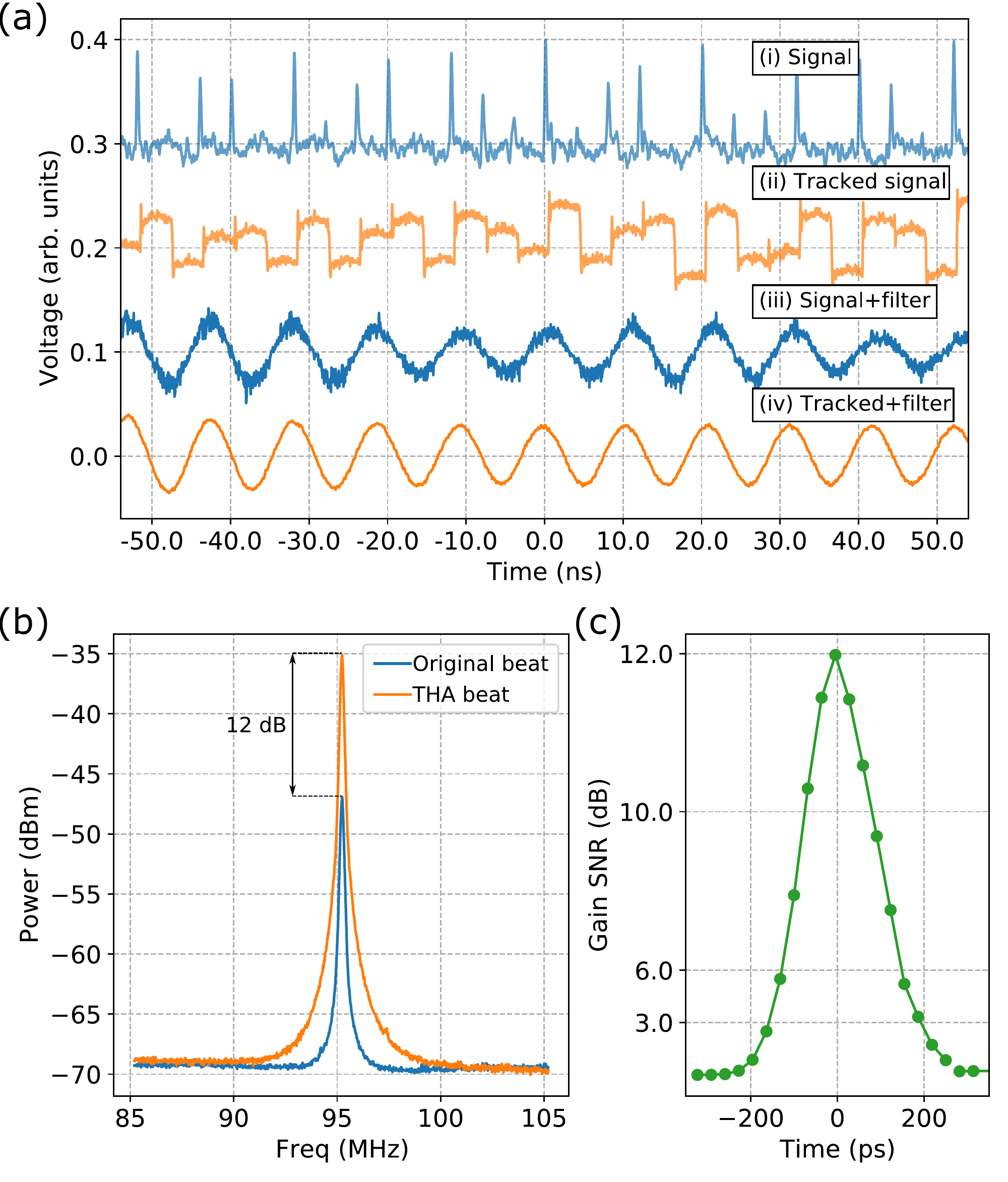}
    \caption{(a) Time-domain illustration of the gated detection technique: trace (i) shows the fast photodiode signal; (ii) shows  the output of the THA. (iii) and (iv) show the traces obtained by band-pass filtering (i) and (ii), respectively. The vertical scale of each signal has been adjusted for easier visualization. All traces were recorded with a fast oscilloscope at \SI{25}{GS/s}. (b) Frequency domain comparison of the results obtained with the conventional heterodyne technique and THA-based gating (RF spectra have been recorded with a 100 kHz resolution bandwidth). The THA beat shows an improvement of 12 dB with respect to the original beat. (c) SNR improvement as a function of the relative phase between the photodiode signal and the clock reference.}
    \label{fig:trace_pd_tha}
\end{figure}

In our laboratory, the optical comb enables the spectral transfer between 
the sub-harmonic wavelengths of our $^{171}$Yb and $^{88}$Sr lattice clocks \cite{Pizzocaro_2020,Barbiero2020} at \SI{1156}{nm} and \SI{1396}{nm} and  a \SI{1542}{nm} ultrastable radiation that is distributed between distant  metrological institutes in Europe using  optical fibers, to enable the comparison of remote optical clocks \cite{Lisdat2016,Schioppo2022,Clivati:20}. To this purpose, we exploit a broadband  branch of the comb that covers the 1000-\SI{2000}{nm} region,  obtained by amplification of the main comb output at \SI{1550}{nm} in an Erbium-doped fiber amplifier followed by spectral broadening in a highly-nonlinear fiber. Figure \ref{fig:optic_setup} (a) shows the measured spectrum of the broad-band comb branch, together with the spectra of our \SI{1156}{nm} and \SI{1542}{nm} lasers. From the measured comb power in the resolution bandwidth of \SI{0.05}{nm} and the modes separation of \SI{250}{MHz}, we estimate that the comb's power per mode is about \SI{12}{nW} at both wavelengths. Nevertheless, power drops of up to \SI{20}{dB} are observed in other regions. Hence, it is difficult to predict the available power at a specific wavelength, and the  comb spectrum cannot be always optimized for cycle-slip-free beatnote detection in different spectral regions simultaneously. \\
We set up the scheme reported in Fig. \ref{fig:optic_setup} (b) to obtain a beatnote between the \SI{1156}{nm} laser and the frequency comb. The two beams are combined in a polarizing beam splitter, and hit a holographic grating that disperses the comb spectrum through a free-space arm of about \SI{1}{m} length, reducing the amount of unwanted radiation hitting the photodetector. Two waveplates are used to  balance their power and  optimize the SNR, and the beatnote is detected with a \SI{2}{GHz} bandwidth InGaAs photodiode. \\
In the present configuration, the beatnote between the \SI{1156}{nm} laser and the comb can be detected with a SNR of about \SI{35}{dB} in a bandwidth of \SI{100}{kHz}, and a similar result is obtained at \SI{1542}{nm}. 
To determine the tolerated amount of comb power reduction before the arising of cycle slips, we reproduce different SNR conditions by attenuating it in the range \SI{2}{dB} - \SI{6}{dB} with neutral density filters. For each configuration we experimentally detect the rate of cycle slips by  redundantly counting the beatnote between our \SI{1156}{nm} laser and the comb, after band-pass filtering it in a bandwidth $B$ of \SI{5}{MHz}. A cycle slip is detected when the difference between the two counted frequencies exceeds the threshold of \SI{0.2}{Hz}, chosen according to the fact that one cycle lost in a measurement interval of \SI{1}{s} causes a jump of \SI{1}{Hz}. Although non critical, the exact threshold value is set as stringent as possible to confidently spot all cycles slips without removing frequency values featuring statistical fluctuations.   Although the rate of cycle slips as a function of the SNR inside B could in principle be computed as  $r = (\frac{1}{2} \text{B})\text{erfc}( 10^{\text{SNR}/20}/\sqrt{2})$, with $\text{erfc}$ being the complementary error function and SNR expressed in dB \cite{sinclair}, the prediction may fail from a practical point of view in an analog-based beatnote processing chain, given the  difficulty in determining the SNR and filter bandwidth with sufficient precision and the strong dependence of $r$ on these parameters. This is highlighted in Figure \ref{fig:snr_cycle}, showing the measured cycle-slips rate (blue dots) at different SNR, as well as the predicted value (thick solid line). The observed discrepancy is attributed to possible systematic biases in our SNR calculations, performed by numerically integrating the noise power as measured on an electrical spectrum analyzer. Fig. \ref{fig:snr_cycle} also indicates the expected cycle slip rate assuming the SNR differs by $+1/-1$ dB (SNR$_1$ and SNR$_{-1}$ respectively). A good qualitative agreement is found in the former case. 
Incidentally, we note that in fact, this aspect is at the heart of cycle-slips flagging methods based on redundant counting, where even nominally identical chains always feature tiny differences that result in significantly-unbalanced cycle-slip rates. \\
The shaded region indicates a cycle slip rate higher than 1/hour: for confident frequency counting at \SI{1e-18}{} the SNR in a \SI{5}{MHz} bandwidth must hence be higher than \SI{15}{dB}. While this condition is fulfilled in our measurement setup, thanks to an adequate comb power per tooth in the region of interest, the strong dependence of the cycle slip rate on the SNR makes the counting extremely sensitive to tiny changes in the beam alignment or polarization due to varying environment conditions. Portions of the spectrum featuring \SI{3}{dB} lower power may not guarantee cycle-slip free counting even in optimal conditions.
In such cases, \SI{10}{dB} improvement on the SNR would be widely sufficient to guarantee suitable measurement conditions. 
Gated detection can enable such an improvement, and we implemented it using the Hittite HMC1061LC5 dual-rank Track and Hold Amplifier (THA). In a THA, the input signal is either copied to the output (transparent mode) or sampled for a short interval and then stored in a capacitor that holds the value for a longer time, regardless of the behavior of the input (hold mode). The HMC1061LC5 is actually composed by two cascaded THAs, each featuring an aperture time of \SI{130}{ps} in the sampling window and a maximum hold time of \SI{2}{ns}, operated in such a way that the latter samples the output of the former close to the end of the hold period. In this way the maximum hold time can be conveniently extended to \SI{4}{ns}, that coincides to the comb pulses separation. \\    
The clock signal opening and closing the gate window and synchronizing the operation of the two THAs must be aligned to the comb pulses. In our setup, this is provided  by a bench-top signal generator whose internal clock is locked to the same signal referencing the comb and whose phase is tuned to match the gating window to the beatnote pulses.\\
As a first step, we compared the traditional detection with the gated detection setup. 
The beatnote between the \SI{1156}{nm} laser and the comb, at \SI{95}{MHz}, is detected with a \SI{2}{GHz}-bandwidth transimpedence photodiode, amplified and splitted in equal parts. Half of the signal is kept as a reference and sent to a \SI{6}{GHz} oscilloscope with a sampling rate of \SI{25}{Gs/s}. The resulting signal, shown in Fig. \ref{fig:trace_pd_tha}a, trace (i), features \SI{300}{ps} wide pulses with a \SI{4}{ns} spacing. The second arm is sent to the THA, and the resulting trace is shown as (ii), with the square-wave components resulting from the holding feature of the THA. To inspect the \SI{95}{MHz} beatnote component, these signals are both band-pass filtered in a bandwidth of \SI{10}{MHz}. The corresponding traces are shown as (iii) and (iv). A considerable improvement of the SNR can be observed at the THA filtered output (iv). This can be attributed to the more favourable duty cycle for this signal as compared to the bare photodiode output, which results in a higher signal power once integrated in the filter bandwidth.
The same information can be obtained by looking at the RF spectrum. Fig. \ref{fig:trace_pd_tha} (b) shows the spectrum of the original beat note (blue) and the one obtained with the THA (orange). The tracked beatnote shows a \SI{12}{dB} increase in the signal power with no change to the noise level. Fig. \ref{fig:trace_pd_tha} (c) shows the SNR improvement as a function of the relative delay between the photodiode signal and the clock reference of the THA, highlighting the importance of aligning the two within a few tens of picoseconds (i.e. a few mm in terms of electric path). 
The maximum SNR improvement is in agreement with the expected value of $2 b/f_\text{rep}$ where $b$ is the photodiode bandwidth and $f_\text{rep}$ the comb's repetition rate.
We also note that the optimal improvement is guaranteed as long as the aperture window of the THA is shorter or has approximately the same duration as the comb pulses. This condition is met in the present system  where the beantote pulse width is \SI{300}{ps}, limited by the  \SI{2}{GHz} bandwidth of the photodiode.

The measurements reported in Fig. \ref{fig:snr_cycle} were then repeated by redundantly counting the tracked beatnote, and under this scheme we were able to achieve cycle-slip-free counting for all the shown measurement conditions.

\begin{figure}
\centering\includegraphics[width=0.8\columnwidth]{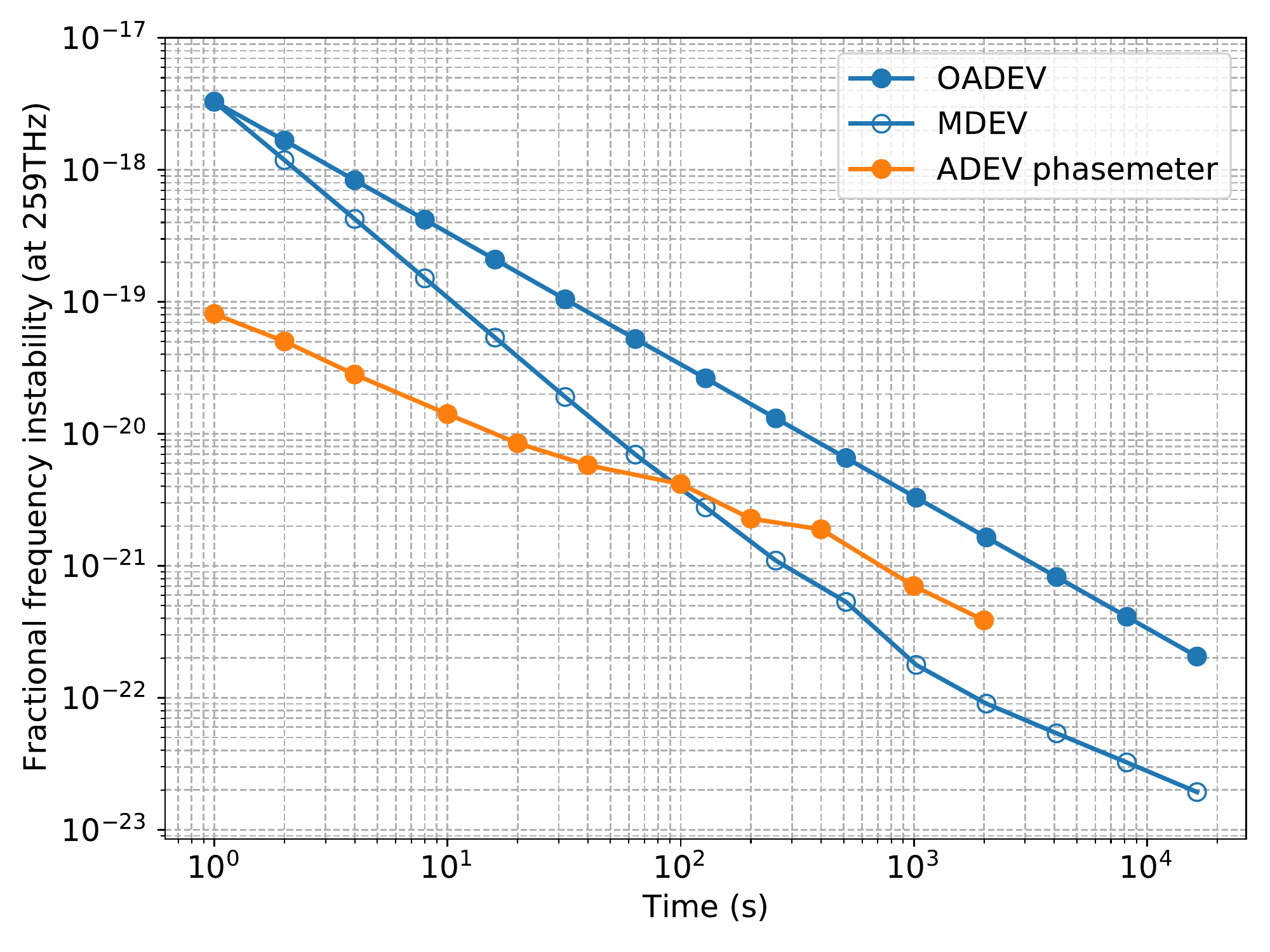}
\caption{In blue the overlapping (full circle) and modified (empty circles) Allan deviation of the difference between a conventional and tracked beatnote, obtained with a multi-channel frequency counter with sampling rate of \SI{1}{kHz} and averaging interval of \SI{1}{s} (an equivalent measurement bandwidth of \SI{0.5}{Hz}). The orange trace is  performed with a digital aliasing-free phasemeter, on the same measurement bandwidth.}
\label{fig:diff_tha_allan}
\end{figure}

From a metrological point of view, it is important to assess the absence of other  noise processes introduced by the gated detection, that may result in additional frequency instability or bias on the counted beatnote. To confirm this, we performed a long-term comparison of the beatnotes obtained with the traditional and gated approach. To ensure a suitably low rate of cycles slips on the untracked beatnote, we adopted a narrowband, \SI{5}{MHz} bandwidth filter and removed the comb attenuation.
In this configuration, the tracked beatnote power improves by \SI{12}{dB} as expected, but the noise also raises above the floor by \SI{4}{dB}, resulting in a net SNR gain of \SI{8}{dB} only. We attribute this effect to an amplification of the comb shot noise, that is normally below the photodiode and the cw laser shot noise, but is selectively amplified by the THA because it is synchronous to the comb pulses \cite{Quinlan2013}. This is confirmed by noting that the noise of the THA output drops to the original level when the comb beam is blocked.
Figure \ref{fig:diff_tha_allan} shows the overlapping (blue full circle) and modified (blue empty circle) Allan deviation of the difference between the direct and tracked beatnotes, obtained with a multi-channel synchronous frequency counter with a native sampling rate of \SI{1}{kHz}. The counter is set in averaging mode with output rate of \SI{1}{Hz} (equivalent measurement bandwidth of \SI{0.5}{Hz}). Owing to a stronger rejection of the measurement jitter, the modified Allan deviation achieves a higher resolution and excludes the presence of major long-term effects and frequency bias at the \SI{2e-23}{} level and confirms the suitability of the THA for high-precision frequency counting. 
It is also interesting to note that the short-term measurement uncertainty obtained with the frequency counter is related to the SNR of the beatnotes, although significantly higher than the predicted limit. Considering that the untracked beatnote features \SI{84}{dB/Hz}, hence contributing a white phase noise $S_0(f)\sim$ \SI{5e-9}{\square{rad}/Hz}, one would expect an overlapping Allan deviation of \SI{5e-20}{}$/\tau$ on a measurement bandwidth of \SI{0.5}{Hz}. This value is close to what we measure with a digital phasemeter with selectable  measurement bandwidth (orange trace), considering that other noise processes not related to the SNR contribute to the measurement noise floor at longer averaging time. Instead, the counter suffers from aliasing due to the limited sampling rate ($f_{s}$) as compared to the signal bandwidth $B/2$, which results in a factor $B/f_{s}$ increase in the white noise power and a corresponding Allan deviation of \SI{3e-18}{}$/\tau$. \\
From these considerations, it is clear that besides  a more robust and cycle-slip free counting, the gated detection could improve the measurement resolution in the typical experimental conditions where frequency counters are widely used. In our experiment, where the the SNR gain offered by the gated detection is \SI{12}{dB} (or a factor 16), the measurement resolution in the computation of optical ratios can be improved by a factor of 4. Such an improvement is significant in measurement conditions with lower SNR, where the aliased detection noise would otherwise prevent measuring at \SI{1e-17}{} resolution.  

\section*{\label{sec:conclusions}Conclusions}
We have developed a real time system based on a THA that improves the SNR of beatnotes between a cw laser and a comb  by up to \SI{12}{dB}, although further improvement could be obtained with faster photodiodes. This gain is already sufficient to  achieve cycle slip free counting in typical experimental conditions, when the comb power is close to the minimum threshold for proper detection, enabling optical frequency ratios and spectral transfer to be reliably performed in a single-branch approach throughout the 1000-\SI{2000}{nm} spectrum and improving the resolution of frequency counter-based measurement to the low \SI{1e-18}{} level. Although in our case the gating signal was obtained with a bulky signal generator, the setup could be further simplified by using commercial Direct Digital Synthetizers, whose phase can be numerically tuned with up to \SI{3}{mrad} resolution \cite{analog_devices_dds}.  Our approach reduces substantially the complexity of the gating detection, because all the signal processing happens in a single chip, without delay lines and additional components. These features make the system versatile, suitable for long-term measurements and easily integrated into conventional optical and electronic setups.

\section*{Acknowledgments}
This project was supported by the European Metrology Program for Innovation and Research (EMPIR) project 20FUN08 NEXTLASERS. The EMPIR initiative is co-funded by the European Union’s Horizon 2020 research and innovation programme and the EMPIR Participating States.

\bibliographystyle{ieeetr}
\bibliography{references.bib}

\end{document}